# Effect of the addition of $Al_2O_3$, $TiO_2$ and ZnO on the thermal, structural and luminescence properties of $Er^{3+}$-doped phosphate glasses

Pablo Lopez-Iscoa[a], Laeticia Petit[b,c], Jonathan Massera[c], Davide Janner[a], Nadia G. Boetti[d], Diego Pugliese[a], Sonia Fiorilli[a], Chiara Novara[a], Fabrizio Giorgis[a], and Daniel Milanese[a,e*]

[a]*Politecnico di Torino, Dipartimento di Scienza Applicata e Tecnologia (DISAT) and INSTM UdR Torino Politecnico, Corso Duca degli Abruzzi 24, 10129, Torino, Italy*

[b]*Optoelectronics Research Centre, Tampere University of Technology, Korkeakoulunkatu 3, 33720, Tampere, Finland*

[c]*Department of Electronics and Communications Engineering, Tampere University of Technology, Korkeakoulunkatu 3, 33720, Tampere, Finland*

[d]*Istituto Superiore Mario Boella, Via P. C. Boggio 61, 10134, Torino, Italy*

[e]*IFN - CNR, CSMFO Lab., Via alla Cascata 56/C, 38123 Povo (TN), Italy*

*** Corresponding author:** Daniel Milanese: daniel.milanese@polito.it

*** Permanent address:** Politecnico di Torino, Dipartimento di Scienza Applicata e Tecnologia (DISAT) and INSTM UdR Torino Politecnico, Corso Duca degli Abruzzi 24, 10129, Torino, Italy

**Abstract**

Er-doped phosphate glasses were fabricated by melt-quenching technique. The changes in their thermal, structural and luminescence properties with the addition of $Al_2O_3$, $TiO_2$ or ZnO were studied. Physical and thermal properties were investigated through density measurement and differential thermal analysis. Structural characterization was performed using the Raman and Infrared spectroscopy. In order to study the influence of the composition on the luminescence properties of the glasses, the refractive index, the luminescence spectra and the lifetime values were measured.

The results show that with the addition of $Al_2O_3$ and $TiO_2$ the phosphate network becomes more connected increasing the glass transition temperature, whereas the addition of ZnO does not show significant changes in the optical, thermal and structural properties but it leads to a larger emission cross-section at 1540 nm as compared to the other glasses. As the site of the $Er^{3+}$ is not strongly affected by the change in the glass composition, we think that the emission properties of the glasses depend on the glass structure connectivity, which has an impact on the $Er^{3+}$ ions solubility.

## 1. Introduction

Since the initial discovery of the Bioglass®, with a composition known as 45S5 corresponding to 45.0 wt% $SiO_2$, 24.5 wt% CaO, 24.5 wt% $Na_2O$ and 6.0 wt% $P_2O_5$ [1], the interest in bioglasses for tissue regeneration has increased [2–5]. Besides silicate glasses, phosphate glasses with a $P_2O_5$ content equal to 50 mol% have shown to be bioactive, degradable and suitable for fiber drawing [6–12]. These bioactive glasses form a hydroxyapatite layer, which is capable of bonding to the connective tissue when placed into body fluids [13]. They have been studied in many biomedical applications, especially for use in bone repair and reconstruction [14], as well as for peripheral nerve regeneration because they allow neuronal cells growth along the fiber's axis [9,15]. The glass system employed in this study is based on the composition 50 $P_2O_5$ – 40 SrO – 10 $Na_2O$ (in mol%), whose bioactivity was previously assessed by J. Massera *et al.* [10]. However, the influence of the addition of erbium and metal oxides on the thermal, structural and luminescence properties of this glass system has never been reported before.

Phosphate glasses are of interest for the engineering of photonic devices, thanks to the following properties: easy processing, good thermal stability and excellent optical characteristics, such as high transparency in the UV-Visible-Near Infrared (UV-Vis-NIR) region [16–20]. Besides, phosphate glasses allow high rare earth (RE) ions solubility. Thus, quenching phenomenon does only occur at very high concentrations of RE ions [20,21]. Due to these properties, phosphate glasses have recently become appealing for optical communications [22] and laser sources as well as optical amplifiers [20,21,23–25]. However, to the best of our knowledge, up to now only few studies focusing on glass compositions that combine both biocompatibility and suitable optical properties have been reported [8,26,27].

The dopant environment around the RE ions plays a significant role in the RE-doped glasses. Specifically, parameters such as the covalency, mass and charge of the ligand atoms affect the luminescence properties as well as the solubility of RE ions in glassy hosts [28]. By adding different metal oxides such as $Al_2O_3$, $TiO_2$ and ZnO, phosphate based glasses are able to modify their structural network, thus changing the glass chemical durability, biocompatibility and other properties [29–33]. In this study, the dopants were selected on the basis of their high covalency and their ability to change the structure of the phosphate glass matrix. In particular, $Al_2O_3$ and $TiO_2$ are known to create cross-linking bonds such as Al-O-P [29,34] and Ti-O-P [35], respectively, while ZnO is considered as a modifier and is responsible for the depolymerization of the glass network [36].

In this paper, we report on the effect of the addition of $Al_2O_3$, $TiO_2$ and ZnO on the thermal, structural and luminescence properties of erbium-doped phosphate glasses.

## 2. Experimental

### 2.1 Glass preparation

Glasses with the compositions in mol% $(0.5\ P_2O_5 - 0.4\ SrO - 0.1\ Na_2O)_{100-x} - (TiO_2/Al_2O_3/ZnO)_x$, with x = 0 and x = 1.5 mol%, were prepared. A fixed amount of $Er_2O_3$, 0.25 mol%, was added to the 100 mol% composition for all the glasses manufactured. The glasses with 1.5 mol% $Al_2O_3$, $TiO_2$ and ZnO were labeled as AlG, TiG and ZnG, respectively, while the glass with x = 0 was labeled as RefG. The glasses were prepared by the conventional melt-quenching technique using $NaPO_3$ (Alfa Aesar), $SrCO_3$ (Sigma-Aldrich, ≥99.9%), $Er_2O_3$ (MV Laboratories Inc., 99.999%), $Al_2O_3$ (Sigma-Aldrich, ≥99.5% α-phase), $TiO_2$ (Sigma-Aldrich, 99.99% rutile) and ZnO (Sigma-Aldrich, ≥99%). $Sr(PO_3)_2$ precursor was independently prepared using $SrCO_3$ and $(NH_4)_2HPO_4$ as raw materials and with a heating up to 850 °C. The chemicals were ground and mixed to prepare a 40 g batch, then placed in a quartz crucible and heated up to 1100 °C for 30 min with a heating rate of 10 °C/min. The melt was poured into a preheated brass mold and annealed at 400 °C for 5 h to decrease the residual stress. Finally, the glasses were cooled down to room temperature. All the glasses were cut and optically polished or ground, depending on the characterization technique.

## 2.2 Physical and thermal properties

The density of the glasses was measured using Archimedes' method with an accuracy of ± 0.02 g/cm$^3$, using distilled water as immersion liquid.

The glass transition temperature ($T_g$) and crystallization temperature ($T_p$) were measured by differential thermal analysis (DTA) using a Netszch JUPITER F1 instrument. The measurement was carried out in a Pt crucible at a heating rate of 10 °C/min. $T_g$ was determined as the inflection point of the endotherm obtained by taking the first derivative of the DTA curve, while $T_p$ was taken as the maximum peak of the exotherm. All measurements were performed with an error of ± 3 °C.

## 2.3 Structural properties

The structural properties of the glasses were assessed using Fourier Transform Infrared (FTIR) Spectroscopy, both in transmission mode and Attenuated Total Reflection mode (FTIR-ATR), and Raman spectroscopy. FTIR-ATR spectra were acquired on glass powders with a Bruker Tensor 27 spectrometer equipped with a liquid nitrogen-cooled mercury–cadmium–telluride (MCT) detector, operating at 2 cm$^{-1}$ resolution, equipped with an ATR cell. The spectra were recorded in the range from 600 to 1400 cm$^{-1}$ and were normalized to the band with maximum intensity (~ 880 cm$^{-1}$).

Raman spectra were acquired with a Renishaw inVia Reflex micro-Raman spectrophotometer (Renishaw plc, Wotton-under-Edge, UK) equipped with a cooled CCD camera using a 785 nm excitation line. The spectra were recorded in the range 600-1400 $cm^{-1}$ and were normalized at the maximum point (~ 1170 $cm^{-1}$).

Semi-quantitative analysis was carried out by using a Scanning Electron Microscope (FESEM, Zeiss Merlin 4248) equipped with an Oxford Instruments X-ACT detector and Energy Dispersive Spectroscopy Systems (EDS/EDX) in order to determine the final composition of the glasses. The composition of all the glasses was found to be in agreement with the nominal one, within the accuracy of the measurement (± 1.5 mol%). Despite the use of quartz crucibles, no Si was found in the EDS analysis of the investigated glasses.

## 2.4 Optical and luminescence properties

The refractive index (n) of the glasses was measured at 5 different wavelengths (633, 825, 1061, 1312 and 1533 nm) by prism-coupling technique (Metricon, model 2010). Ten scans were performed for each wavelength. Estimated error of the measurement was ± 0.001. The experimental data were fitted using Sellmeier's equation:

$$n^2(\lambda) = 1 + \frac{B_1 \cdot \lambda^2}{\lambda^2 - C_1} + \frac{B_2 \cdot \lambda^2}{\lambda^2 - C_2} + \frac{B_3 \cdot \lambda^2}{\lambda^2 - C_3} \quad (1)$$

where $\lambda$ is the wavelength and $B_{1,2,3}$ and $C_{1,2,3}$ are the experimentally determined Sellmeier's coefficients.

The absorption spectra in the range 2500-4000 $cm^{-1}$ were recorded by means of a FTIR spectrometer (Alpha, Bruker Optics, Ettlingen, Germany) working in transmission mode and equipped with a DTGS detector. The measurements were performed at room temperature and corrected for Fresnel losses and glass thickness.

The UV-Vis absorption spectra were measured at room temperature from 190 to 1600 nm using an UV-Vis-NIR Agilent Cary 5000 spectrophotometer (Agilent, Santa Clara, CA, USA). The absorption cross-section ($\sigma_{Abs}$) was calculated from the experimentally measured absorption coefficient and from $Er^{3+}$ ions concentration in the glass, using the following formula:

$$\sigma_{Abs}(\lambda) = \frac{2.303}{NL}\log\left(\frac{I_0}{I}\right) \quad (2)$$

where $\log(I_0/I)$ is the absorbance, $L$ is the thickness of the sample (in cm) and $N$ is the rare-earth ion concentration (ions/cm$^3$). The $Er^{3+}$ ions concentration was calculated from the measured glasses density.

The emission spectra in the 1400-1700 nm range were measured with a Jobin Yvon iHR320 spectrometer equipped with a Hamamatsu P4631-02 detector and a filter (Thorlabs FEL 1500). Emission spectra were obtained at room temperature using an excitation monochromatic source at 976 nm, emitted by a single-mode fiber pigtailed laser diode (CM962UF76P-10R, Oclaro). The glass samples used for the absorption and emission measurements were optically polished disks of 1 mm of thickness.

The emission cross-section ($\sigma_e$) spectrum was calculated from the absorption cross-section spectrum using the McCumber's equation [37]:

$$\sigma_e = \sigma_{Abs} \cdot \exp\frac{(\varepsilon - E)}{kT} \quad (3)$$

where $\sigma_{Abs}$ is defined by Eq. (2), $\varepsilon$ is the photon energy at which the two spectra cross at temperature $T$, $E$ is the energy in eV and $k$ is the Boltzmann's constant.

The fluorescence lifetime of $Er^{3+}$:$^4I_{13/2}$ energy level was obtained by exciting the samples with a fiber pigtailed laser diode operating at the wavelength of 976 nm, recording the signal using a digital oscilloscope (Tektronix TDS350) and fitting the decay traces by single exponential. Estimated error of the measurement was ± 0.20 ms. The detector used for this measurement was a Thorlabs PDA10CS-EC.

## 3. Results

The physical and thermal properties of the investigated glasses are reported in Table 1. The addition of $Al_2O_3$ (AlG), $TiO_2$ (TiG) or ZnO (ZnG) has no impact on the density. However, the addition of $Al_2O_3$ and $TiO_2$ increases the $T_g$, while the addition of ZnO does not modify the $T_g$. Table 1 also shows $\Delta T$, the temperature difference between $T_p$ and $T_g$, which is an indicator of the glass resistance to crystallization. It is interesting to point out that AlG has the highest $\Delta T$ as compared to the other glasses. All the investigated glasses show $\Delta T$ higher than 100 °C.

The FTIR-ATR and Raman spectra of the glasses are shown in Fig. 1a and Fig. 2a, respectively. All the spectra were normalized to the band with maximum intensity, thus all the discussed intensity changes are expressed relatively to the main peak.

The FTIR-ATR spectra exhibit a broad band between 650 and 800 $cm^{-1}$, three absorption bands located at around 1250, 1085 and 880 $cm^{-1}$, and two shoulders at ~ 1160 and 980 $cm^{-1}$. The addition of ZnO leads to very small changes in the IR spectra: the intensity of the shoulder at 980 $cm^{-1}$ and of the band at 1085 $cm^{-1}$ are slightly decreased, as reported in Fig. 1b and Fig. 1c, respectively. Moreover, a small increase in intensity of the band at 1250 $cm^{-1}$ as compared to the band at 880 $cm^{-1}$ is evidenced in Fig. 1d. The addition of $Al_2O_3$ leads to a decrease in the intensity of all the bands as compared to the main band. One can also observe a small shift of the bands position to higher wavenumbers. The addition of $TiO_2$ slightly increases the intensity of the shoulder at 980 $cm^{-1}$ (see Fig. 1b) and of the band at 1250 $cm^{-1}$ (see Fig. 1d) and decreases the intensity of the band in the 650-800 $cm^{-1}$ range and at 1085 $cm^{-1}$ (see Fig. 1c). The position of the main band and of the band at 1250 $cm^{-1}$ is also shifted to higher wavenumbers.

The Raman spectra of the glasses exhibit defined bands at ~ 700, 1170 and 1280 $cm^{-1}$ and several bands between 800 and 1110 $cm^{-1}$. With the addition of ZnO, the band at 700 $cm^{-1}$ is slightly reduced (see Fig. 2b). The band at 1250 $cm^{-1}$ is shifted to higher wavenumbers and becomes narrower. With the addition of $TiO_2$, all the bands remained unchanged in terms of intensity and shape. As for the impact of the $Al_2O_3$ addition, the intensity of the band at 700 $cm^{-1}$ increases and the main band at 1170 $cm^{-1}$ is broader, probably due to an increase of the topological disorder within the glass matrix (see Fig. 2b).

Fig. 3 shows the refractive index values of the glasses measured at 5 different wavelengths fitted with the Sellmeier's formula. RefG, AlG and ZnG show similar refractive index values, while TiG displays higher values at all the wavelengths.

Fig. 4 shows the IR spectra of the glasses registered in absorption mode. The spectra exhibit a broad absorption band between 2700 and 3500 $cm^{-1}$, which corresponds to the stretching vibration mode of $OH^-$ groups in several oxide glasses [38]. The TiG system is characterized by the most intense band, suggesting a larger $OH^-$ population as compared to the other glasses.

The UV-Vis absorption spectra of all the glasses are reported in Fig. 5. The addition of $Al_2O_3$ and ZnO slightly shifts the UV absorption edge to longer wavelengths, and this effect is more pronounced with the addition of $TiO_2$. The spectra exhibit several bands characteristics of the $Er^{3+}$ ion 4f-4f transitions from the ground state to various excited levels [39,40].

The absorption cross-section spectra at around 980 and 1550 nm, calculated using the Eq. (2), are represented in Figs. 6a and 6b, respectively. All the glasses show similar absorption cross-sections within the accuracy of the measurement.

Emission and normalized emission spectra, measured in the wavelength range 1400-1700 nm under excitation at 976 nm, are illustrated in Fig. 7a and 7b, respectively. They exhibit the typical emission band assigned to the $Er^{3+}$ transition from $^4I_{13/2}$ to $^4I_{15/2}$, which changes with the composition. The addition of ZnO and $Al_2O_3$ increases the intensity of the emission at 1550 nm, while the RefG and TiG exhibit emission with similar intensity. As evident in Fig. 7b, the change in the glass composition has no impact on the shape of the emission band. The McCumber's emission cross-section spectra of the glasses, calculated using the Eq. (3), are shown in Fig. 7c and their respective values at 1550 nm are listed in Table 1. The ZnG exhibits the highest emission cross-section, in agreement with the emission measurements.

Luminescence decay curves from the $^4I_{13/2}$ to the $^4I_{15/2}$ emission upon 976 nm excitation are reported in Fig. 8, and their corresponding lifetime values are listed in Table 1. The AlG shows similar lifetime value as compared to the reference glass, whereas the TiG exhibits the lowest lifetime and the ZnG the highest one.

# 4. Discussion

## 4.1 Structural properties

The physical and thermal properties of the investigated glasses are reported in Table 1. The addition of $Al_2O_3$, $TiO_2$ or ZnO has no significant impact on the density, but remarkable effects are reported regarding the thermal properties of the prepared glasses. The addition of ZnO has no impact on $T_g$, it decreases slightly $T_p$ and thus $\Delta T$, whereas the addition of $TiO_2$ and $Al_2O_3$ increases $T_g$, $T_p$ and $\Delta T$. The increase in $T_g$ could indicate that the addition of $TiO_2$ and $Al_2O_3$ improves the strength of the network, whereas Zn is believed to act as a network modifier, in agreement with Schwarz *et al.* [41]. It is worthwhile noting that the AlG exhibits the highest $T_g$, $T_p$ and $\Delta T$, thus suggesting that Al has a higher impact on the bond strength than the other elements [42]. A $\Delta T$ higher than 100 °C for all glasses suggests their reasonable thermal stability.

In the IR spectra shown in Fig. 1a, the broad band between 650 and 800 cm$^{-1}$ may include symmetric vibrational modes $\nu_{sym}$ (P-O-P) of $Q^2$ units [43]. The main band at ~ 880 cm$^{-1}$ is attributed to the asymmetric vibrational mode $\nu_{as}$ (P-O-P) in chains of $Q^2$ units [43–46]. The various bands between 930 and 1010 cm$^{-1}$ are often related to rings-type formation in the glass network [47]. The shoulder centered at ~ 980 cm$^{-1}$ and the band peaked at 1085 cm$^{-1}$ correspond to the symmetric and asymmetric stretching vibrations of $PO_3^{2-}$ in $Q^1$ units, respectively [45,46,48,49]. Besides, the band at 1085 cm$^{-1}$ can be attributed to an overlap between $PO_3^{2-}$ of $Q^1$ units and $PO_2^-$ of $Q^2$ groups in metaphosphate [50]. The shoulder at 1160 cm$^{-1}$ and the absorption band at 1250 cm$^{-1}$ correspond to the symmetric and asymmetric vibrations of $PO_2^-$ in $Q^2$ units, respectively [43,45,46,48,51]. No bands are revealed at wavenumbers higher than 1300 cm$^{-1}$, where the ν(P=O) of $Q^3$ groups typically locate. These IR spectra clearly indicate the presence of a metaphosphate structure, which is confirmed from the analysis of the Raman spectra presented in Fig. 2a. The Raman band at around 700 cm$^{-1}$ corresponds to the symmetric stretching of bridging $\nu_{sym}$(O-P-O) of $Q^2$ groups and the band at 1025 cm$^{-1}$ to the symmetric stretching ν(P-O) of terminal groups ($Q^1$) [43]. The bands at 1170 and 1280 cm$^{-1}$ can be ascribed to the symmetric and asymmetric stretching vibrations of non-bridging ν($PO_2$) of $Q^2$ groups, respectively [52–54]. These spectra are clearly index of the presence of a metaphosphate structure, as suggested by Velli *et al.* [55]. A few amount of terminal groups $Q^1$ can also be observed, while there is no evidence of the presence of $Q^3$ units, usually responsible of Raman shifts higher than 1300 cm$^{-1}$.

The addition of ZnO barely affects the IR and Raman spectra, as corroborated by the analysis of the density and of the thermal properties of this glass. From the small changes in the IR and Raman spectra, Zn seems to act as a modifier, leading to a depolymerization of the phosphate network and to a less cross-linked network. Its concentration in the present glass is probably too small to induce noticeable changes in its physical, thermal and structural properties.

Regarding the TiG, the incorporation of $TiO_2$ has a small influence on the broadening and shifting of the $Q^2$ units bonds. The decrease in intensity of the IR band at 1085 cm$^{-1}$ and the increase in intensity of the IR shoulder at 980 cm$^{-1}$ might be related to the decrease of bridging oxygens on incorporation of $TiO_2$ due to the formation of P-O-Ti bonds, as suggested by Kiani *et al.* [56]. The addition of $TiO_2$ leads to an increase in $T_g$, maybe ascribable to the distortion of the glass network due to the formation of three-dimensional networks of P-O-Ti linkages, as suggested by Segawa *et al.* [57]. The shift of the main Raman band at 1170 $cm^{-1}$ confirms the change in the average length of the P-O-P bond with the addition of $TiO_2$. The concentration of $TiO_2$ is probably too small to display the Raman band assigned to Ti-O bond, which appears at 930 $cm^{-1}$ according to Kiani *et al.* [56]. As shown in Fig. 3, the addition of $TiO_2$ increases the refractive index, probably due to an increase in the electron density. This effect is usually ascribed to the creation of more $Q^1$ units, which have higher polarizability than $Q^2$ units [58,59].

The addition of $Al_2O_3$ has the highest impact on the glass structure. The decrease in the intensity of all the IR bands, as compared to the main band, indicates that $Al^{3+}$ ions are expected to enter gradually the network, as reported by Saddeek *et al.* [60]. The non-bridging oxygens of P=O bonds may be converted into bridging oxygens upon formation of P-O-Al bonds and the links O-Al-O replace O-P-O as aluminum enters the network [60]. This process is associated with the small shift to higher wavenumbers of the IR bands position. These P-O-Al cross-linking bonds between phosphate chains increase the network connectivity, as suspected from the increase in intensity of the Raman band at 700 $cm^{-1}$. This is also in agreement with the increase of $T_g$.

The IR absorption spectra of the glasses between 2400 and 4000 $cm^{-1}$ are shown in Fig. 4. The spectra show a similar band for all glasses, however, the TiG presents the highest absorption intensity for the signal at 2900 $cm^{-1}$, ascribable to a higher $OH^-$ content [38]. This larger $OH^-$ population if compared to that of the other glasses might indicate that the formation of the P-O-Ti bonds distorts the phosphate network, increasing the free volume in the glass followed by an increase of the $OH^-$ absorption.

**4.2 Optical properties**

The absorption spectra of the glasses are shown in Fig. 5. The UV absorption edge was not clearly affected by the addition of $Al_2O_3$ or ZnO. However, with the addition of $TiO_2$, the UV absorption edge shifted to longer wavelengths due to the presence of $Ti^{3+}$, as explained by Zikmund *et al.* [61]. In the 250-500 nm range several bands are observed. These bands are characteristics of the $Er^{3+}$ ion 4f-4f transitions from the ground state to various excited levels. In the inset of Fig. 5, the absorption edge at the UV end of the UV-Vis spectra can be seen in more detail. As stated in [62,63], an increase in the polymerization of the phosphate network leads to a shift of the UV absorption edge to lower wavelength values. Due to the formation of Al-P-O bond, the polymerization in the AlG is more significant than in TiG and ZnG, thus leading to a more pronounced shift of the UV absorption edge to lower wavelengths with respect to the other glasses.

The absorption cross-section values at around 980 and 1550 nm are represented in Fig. 6a and 6b, respectively. The absorption cross-sections were calculated from the absorption coefficients using Eq. (2). Within the error of the measurement, the absorption cross-section remains unchanged with the addition of $Al_2O_3$, $TiO_2$ and ZnO, thus indicating that the site of the $Er^{3+}$ ions is not strongly influenced by the changes in the glass composition. This clearly shows that although Al, Ti and Zn have different impacts on the glass structure, they do not participate to the second coordination shell around the $Er^{3+}$ ions.

**4.3 Luminescence properties**

The emission spectra upon excitation at 976 nm are reported in Fig. 7a. The ZnG exhibits the highest emission at 1540 nm, whereas the TiG and AlG show the lowest emission at 1540 nm as compared to the RefG. The shape of the emission band (see Fig. 7b) is unaffected by the changes in glass composition, thus confirming that $Al_2O_3$, $TiO_2$ and ZnO have no significant impact on the site of the $Er^{3+}$ ions. Additionally, the higher emission of the ZnG as compared to the other glasses cannot be related to the absorption properties of the glasses at 980 nm, as all the investigated glasses possess similar absorption cross-sections at this pump wavelength. It is not possible either to relate it to the $OH^-$ groups, which are known to diminish the emission intensity by non-radiative phenomena [38], as the glasses have also similar $OH^-$ content except for the TiG. Therefore, we believe that the different emission properties of the glasses could be related to the different $Er^{3+}$ ions solubility within the glass matrix and thus to the different network connectivity. Similarly, the larger emission and emission cross-section (see Table 1 and Fig. 7c) of the ZnG could be associated to a better solubility of the $Er^{3+}$ ions into the ZnG due to the depolymerization of the network induced by the addition of ZnO. On the contrary, the addition of $Al_2O_3$ and $TiO_2$, which are suspected to increase the network connectivity, would most probably reduce the Er-Er distance and so the $Er^{3+}$ ions solubility. Further investigations on $Er^{3+}$ solubility in the glasses are currently ongoing in order to clarify the above observation.

The lifetime values of $Er^{3+}$:$^4I_{13/2}$ level in the investigated glasses are shown in Fig. 8 and listed in Table 1. Within the measurement error, all investigated glasses exhibit similar lifetime values, except for the TiG which displays a lower one. This could be related to the slightly higher $OH^-$ content in this glass as compared to the other ones. As shown in Fig. 9, a nearly linear relationship between the reciprocal of the lifetime and the absorption coefficient at 2900 $cm^{-1}$ related to the $OH^-$ content is observed. The decrease in lifetime and emission with increasing the $OH^-$ concentration due to the energy transfer from $Er^{3+}$ ions to quenching centers like $OH^-$ groups is in agreement with Yan *et al.* [38].

## 5. Conclusions

The effect of the glass composition on the physical, structural and luminescence properties of $Er^{3+}$ containing glasses in the system $P_2O_5$ – SrO – $Na_2O$ was investigated. The addition of $Al_2O_3$ and $TiO_2$ has a slight impact on the density, glass structure and on the spectroscopic (absorption and emission) properties of the fabricated glasses. Instead, the addition of ZnO, which is suspected to slightly depolymerize the phosphate network, increases the intensity of the emission at 1540 nm, although this glass has similar absorption cross-section at the pump wavelength compared to the other investigated glasses. As the site of the $Er^{3+}$ ions is not influenced by the change in the glass composition, the modification in the glass connectivity is supposed to affect the $Er^{3+}$ ions solubility and thus the emission properties of the glasses. The investigated glasses possess also a good thermal stability and are therefore promising for the fabrication of fiber lasers and amplifiers.

## Acknowledgements

The research leading to these results has received funding from the European Union's Horizon 2020 research and innovation programme under the Marie Sklodowska-Curie grant agreement No. 642557. We also acknowledge the Academy of Finland for "COMPETITIVE FUNDING TO STRENGTHEN UNIVERSITY RESEARCH PROFILES" funded by Academy of Finland, decision number 310359".

## Conflict of interest

The authors declare that there is no conflict of interest regarding the publication of this article.

**Figure captions**

**Fig. 1.** FTIR-ATR spectra of the glasses (a). The insets show the zoom-up of the bands at 980 (b), 1085 (c) and 1160 (d) $cm^{-1}$, respectively.

**Fig. 2.** Raman spectra of the glasses (a). The insets show the zoom-up of the bands at 700 (b) and 1170 $cm^{-1}$ (c), respectively.

**Fig. 3.** Refractive index values of the prepared glasses at 5 different wavelengths fitted with the Sellmeier's formula. The filled squares represent the experimental data, while the continuous lines are the fitting curves.

**Fig. 4.** IR absorption spectra of the glasses.

**Fig. 5.** UV-Vis absorption spectra of the investigated glasses. The inset shows the UV edge of RefG, AlG and ZnG in the range between 195 and 210 nm.

**Fig. 6.** Absorption cross-section of the investigated glasses centered at 980 (a) and 1500 nm (b).

**Fig. 7.** Emission (a), normalized emission (b) and emission cross-section (c) spectra of the investigated glasses.

**Fig. 8.** Room temperature decay curves of the $^4I_{13/2}$ level of $Er^{3+}$ ions in the glasses obtained under excitation at 980 nm. The intensity data are reported on a Log scale.

**Fig. 9.** Decay rate, defined as the inverse of the $Er^{3+}$:$^4I_{13/2}$ level lifetime, as a function of the absorption coefficient of $OH^-$ vibration band at 2900 $cm^{-1}$ of all the glasses. The experimental data were fitted through the formula reported in [64].

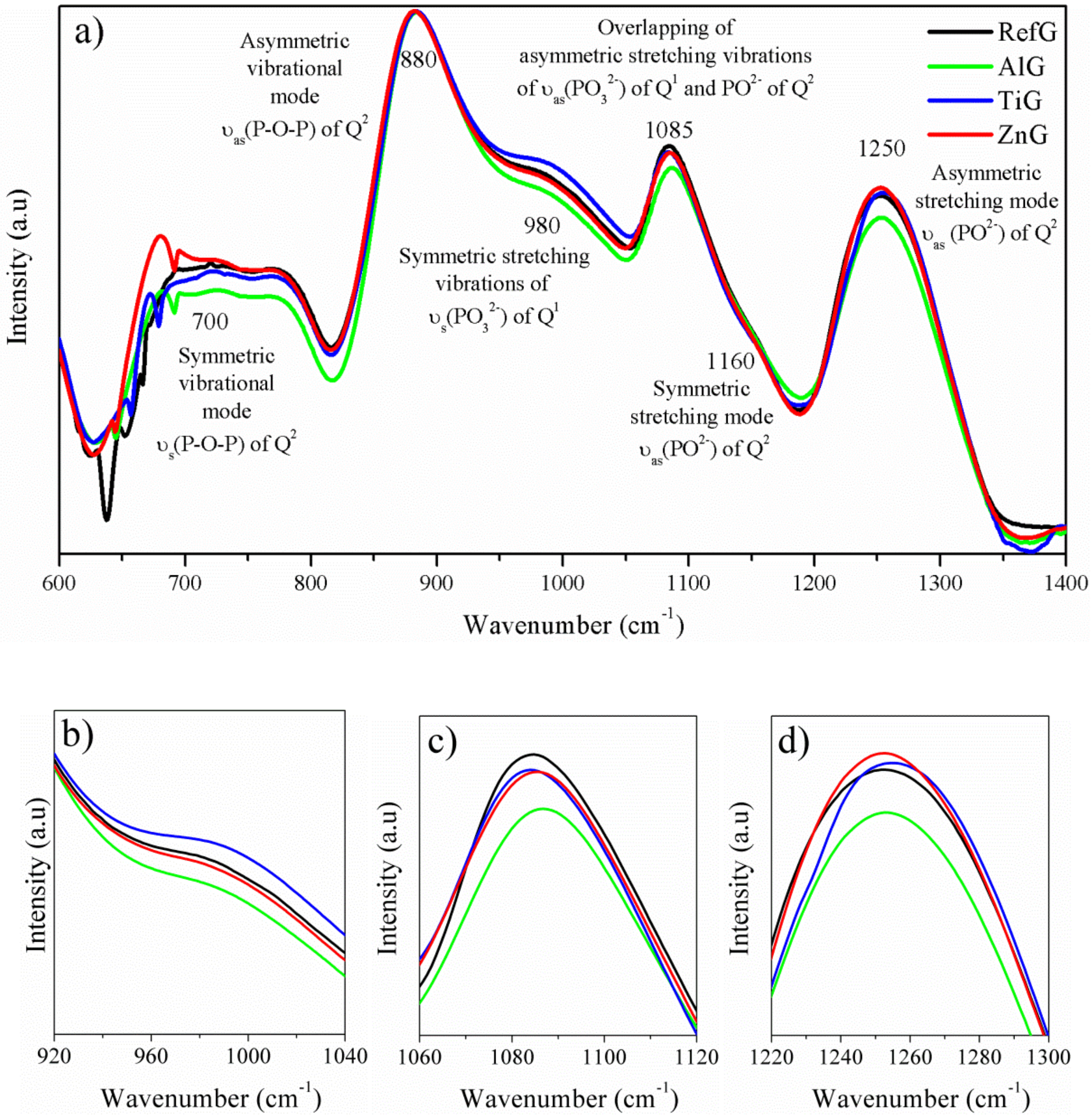
a)
Asymmetric vibrational mode
υas(P-O-P) of Q2
880
Overlapping of asymmetric stretching vibrations of υas(PO3 2-) of Q1 and PO2- of Q2
1085
1250
Asymmetric stretching mode
υas (PO2-) of Q2
RefG
AlG
TiG
ZnG
980
Symmetric stretching vibrations of
υs(PO3 2-) of Q1
700
Symmetric vibrational mode
υs(P-O-P) of Q2
1160
Symmetric stretching mode
υas(PO2-) of Q2
Intensity (a.u)
Wavenumber (cm-1)
b)
c)
d)

**Fig. 1.**

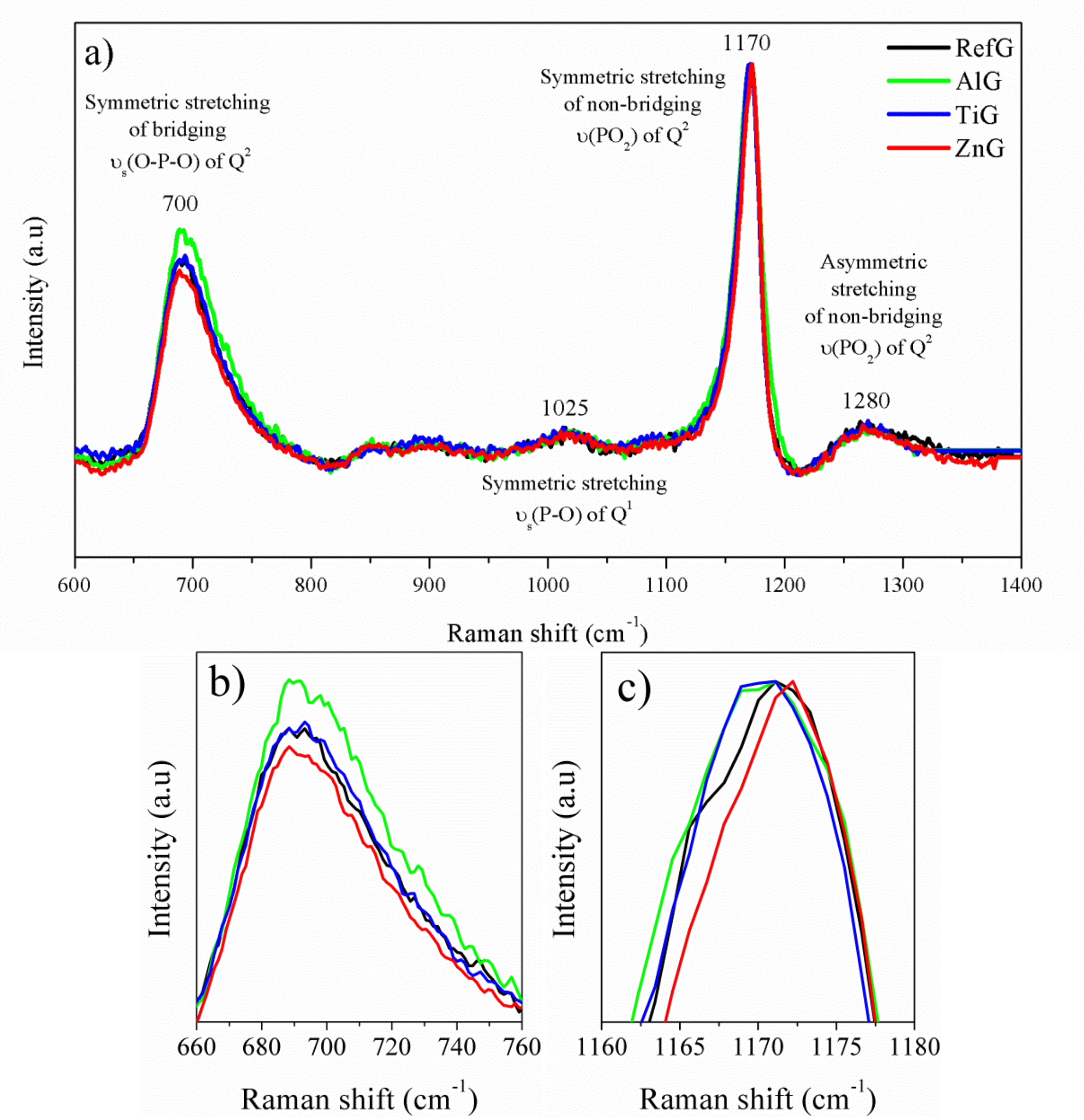

**Fig. 2.**

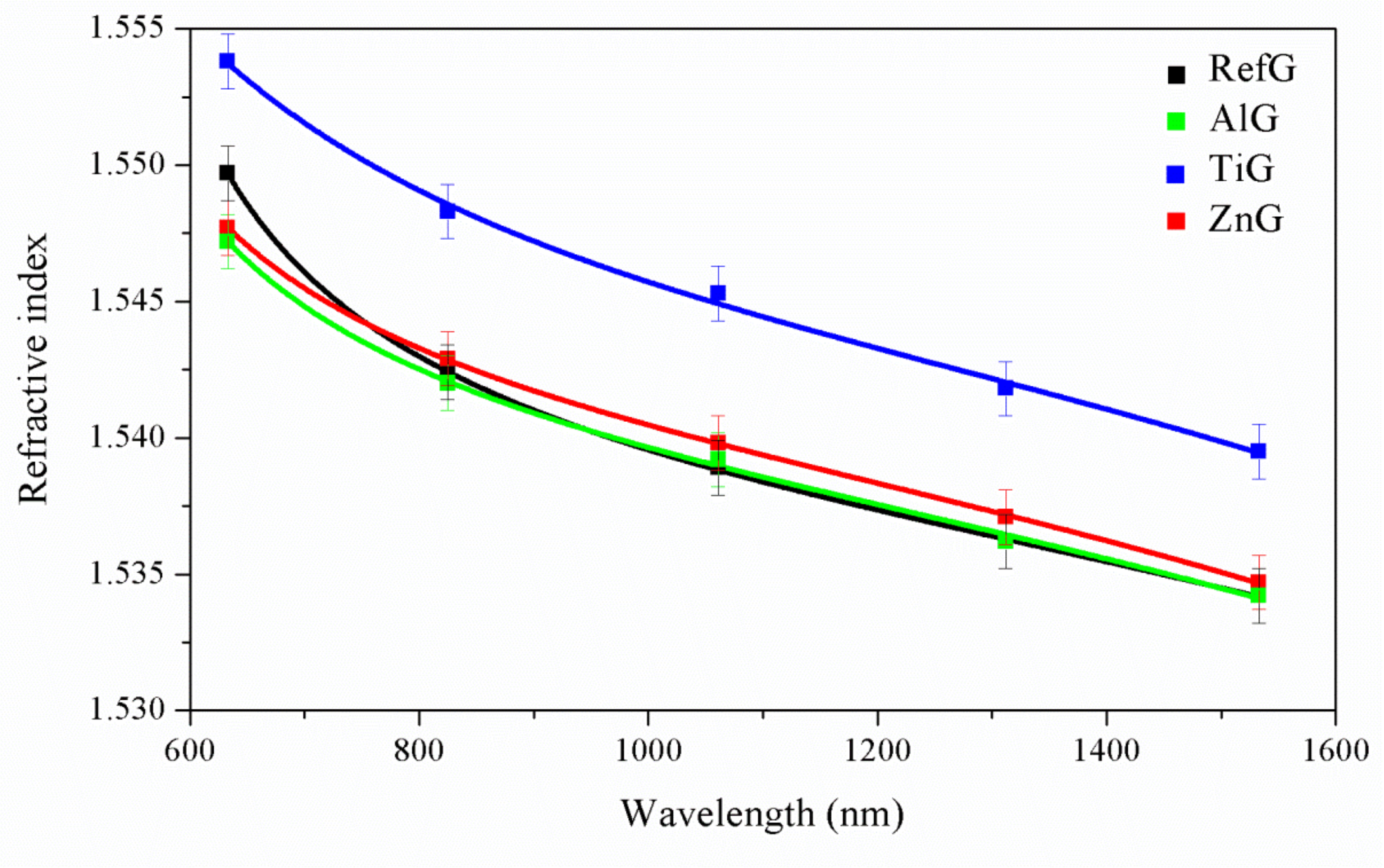

**Fig. 3.**

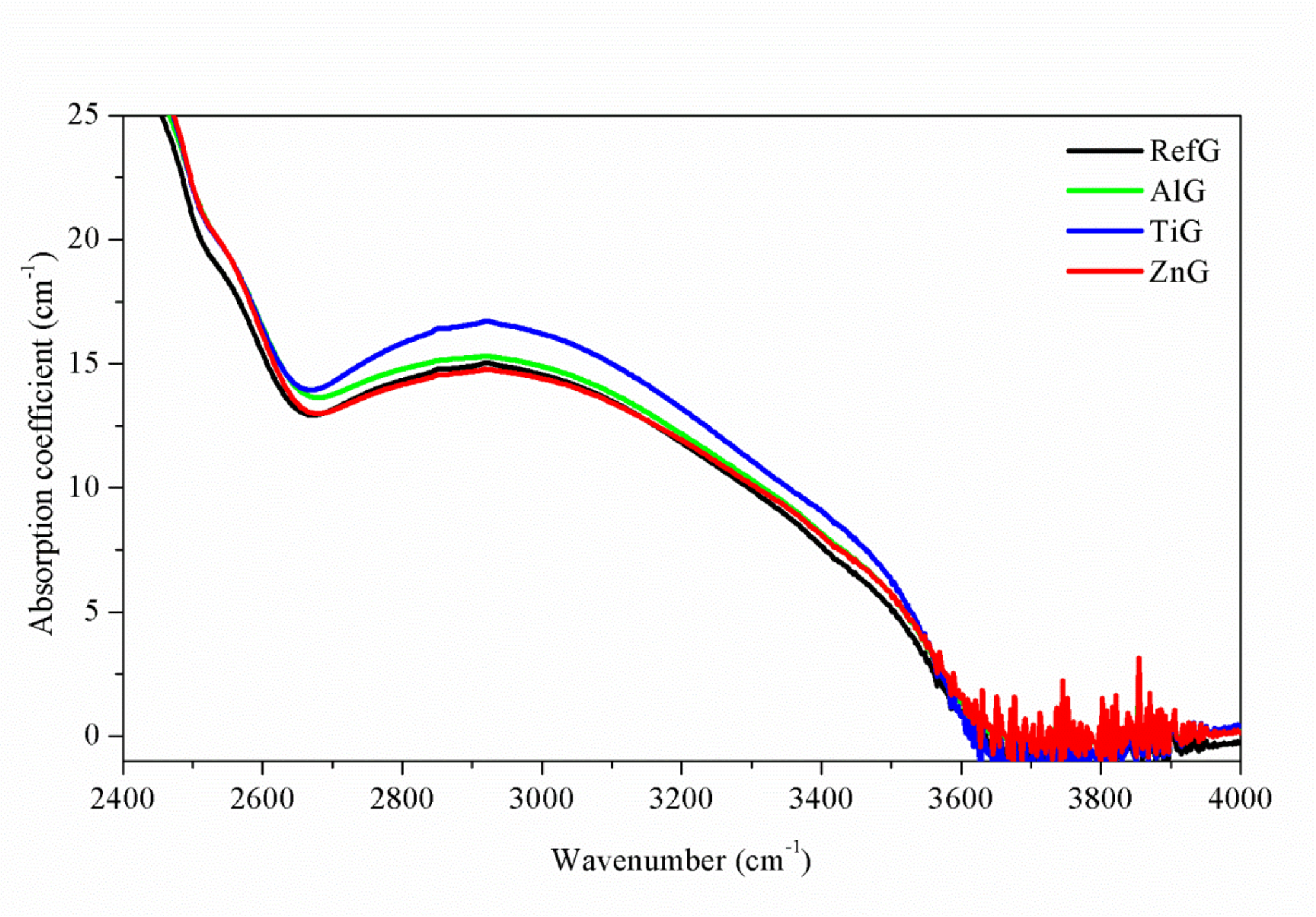

**Fig. 4.**

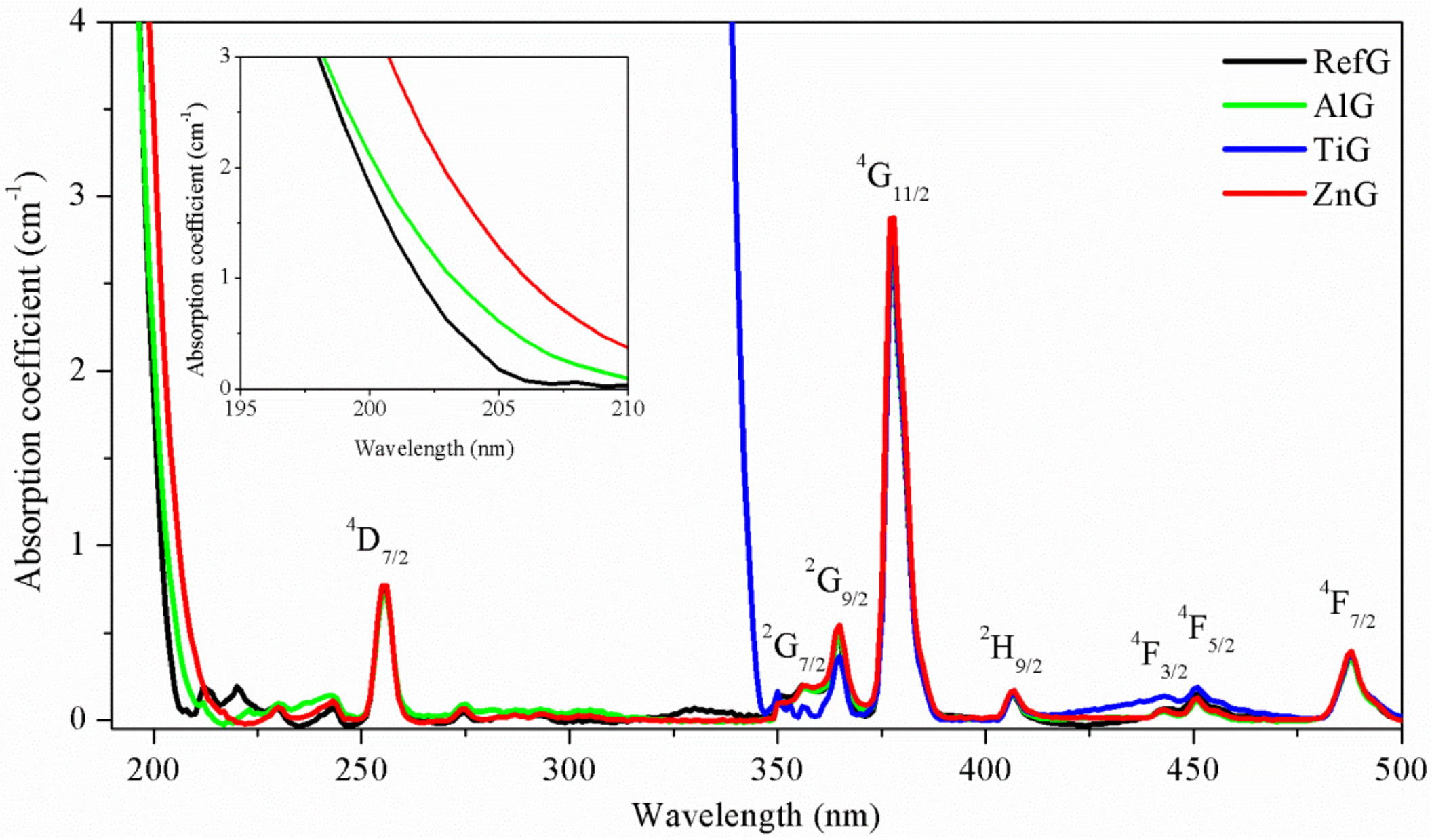

**Fig. 5.**

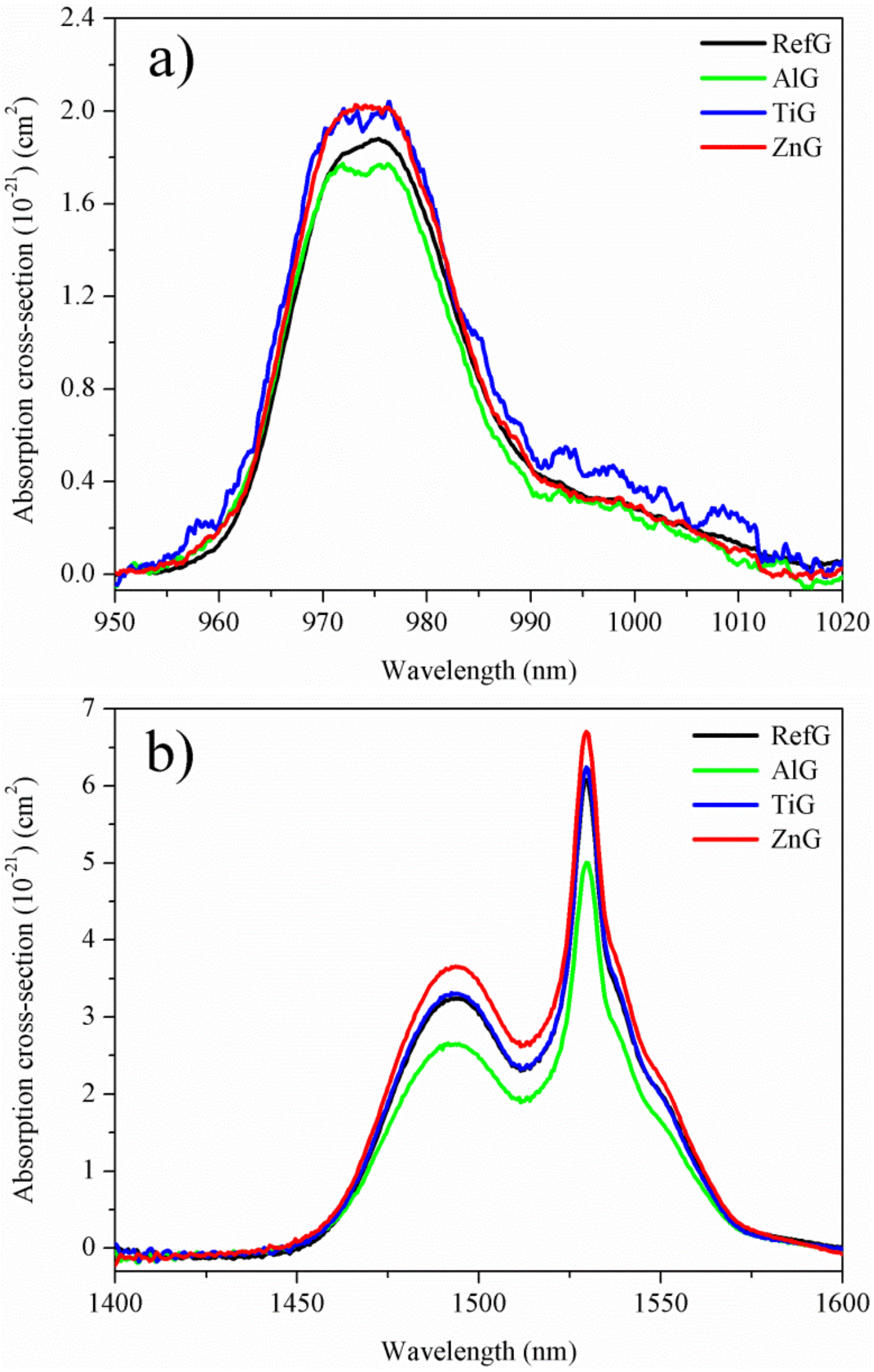

**Fig. 6.**

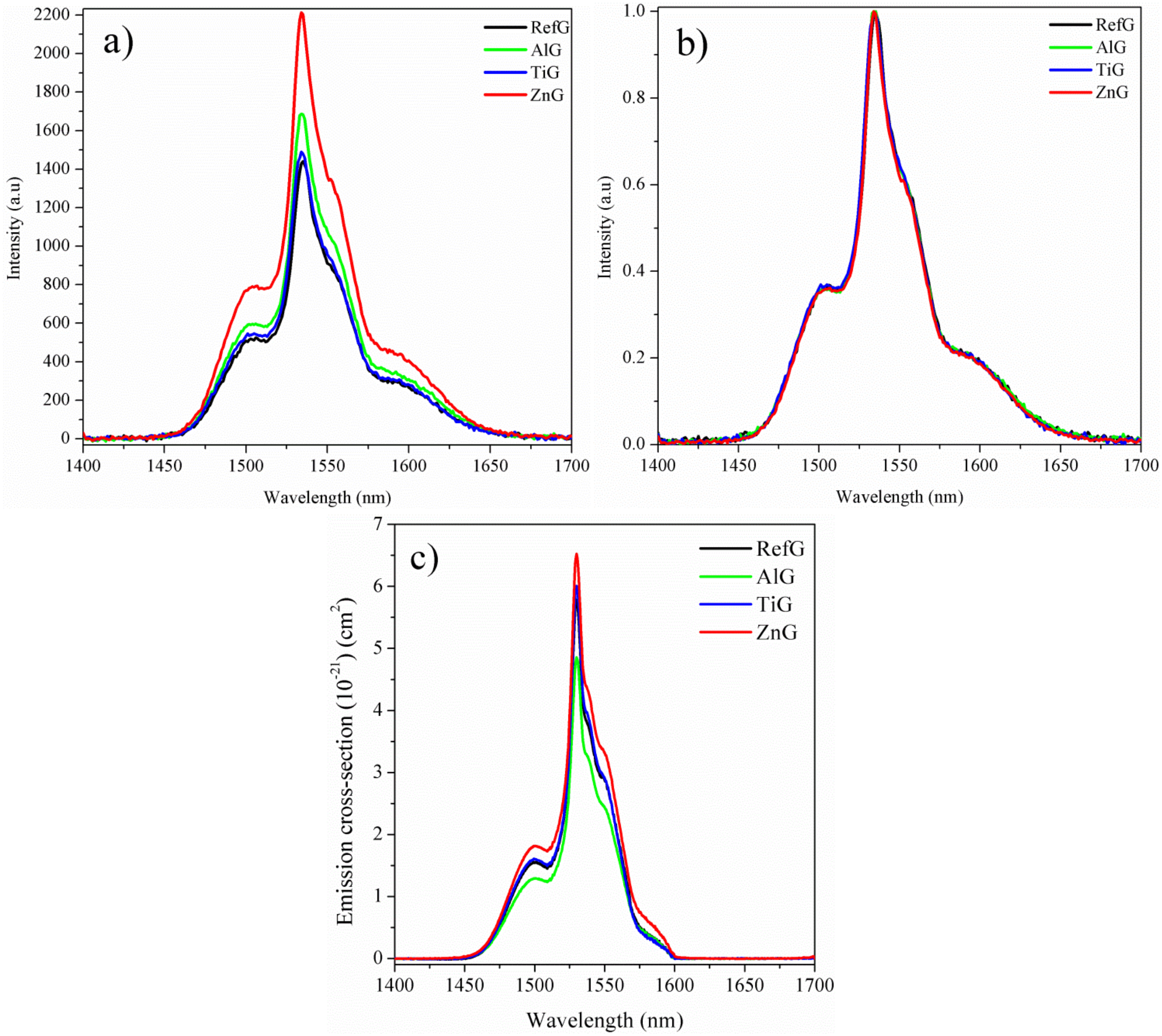

**Fig. 7.**

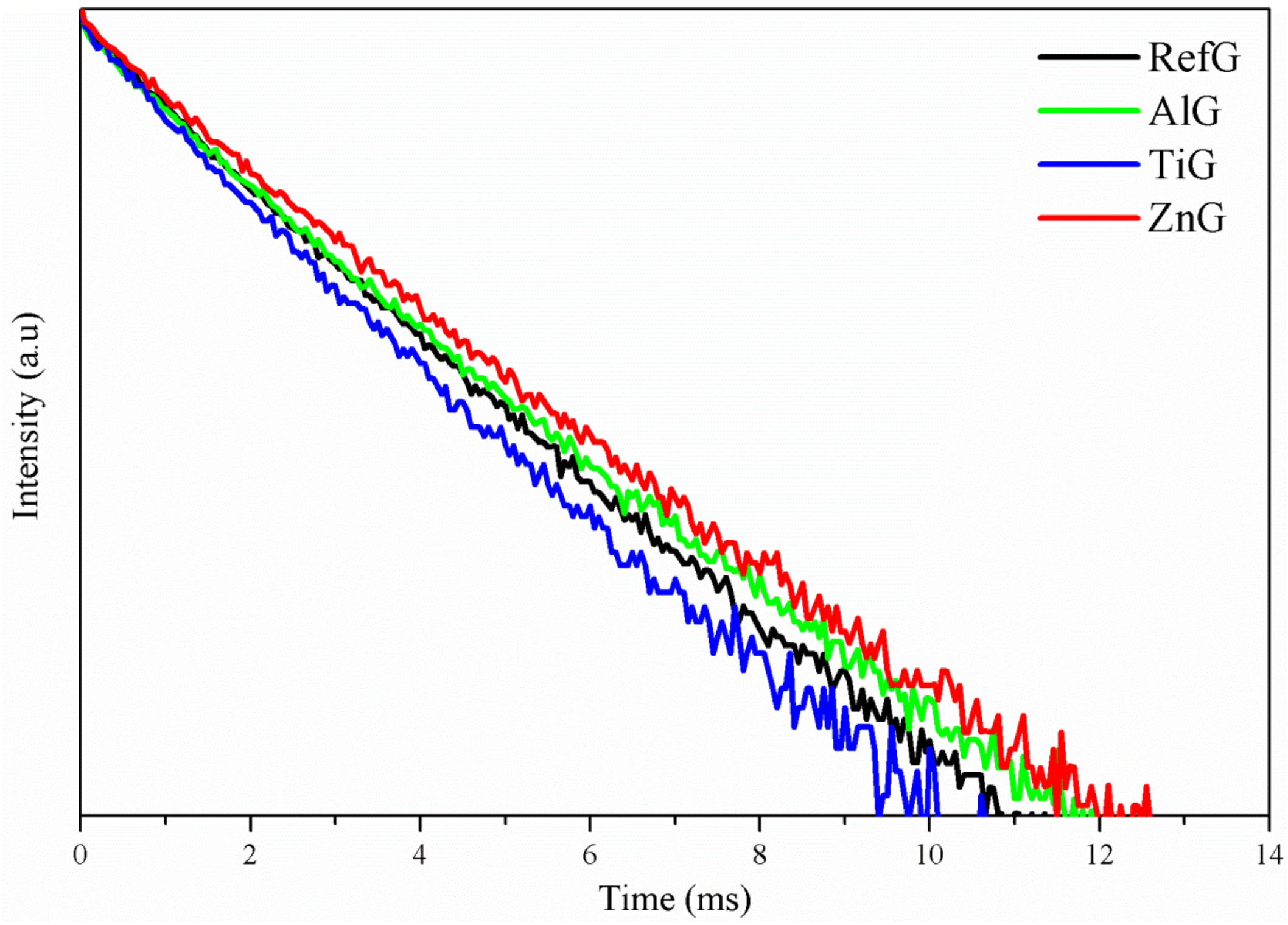

**Fig. 8.**

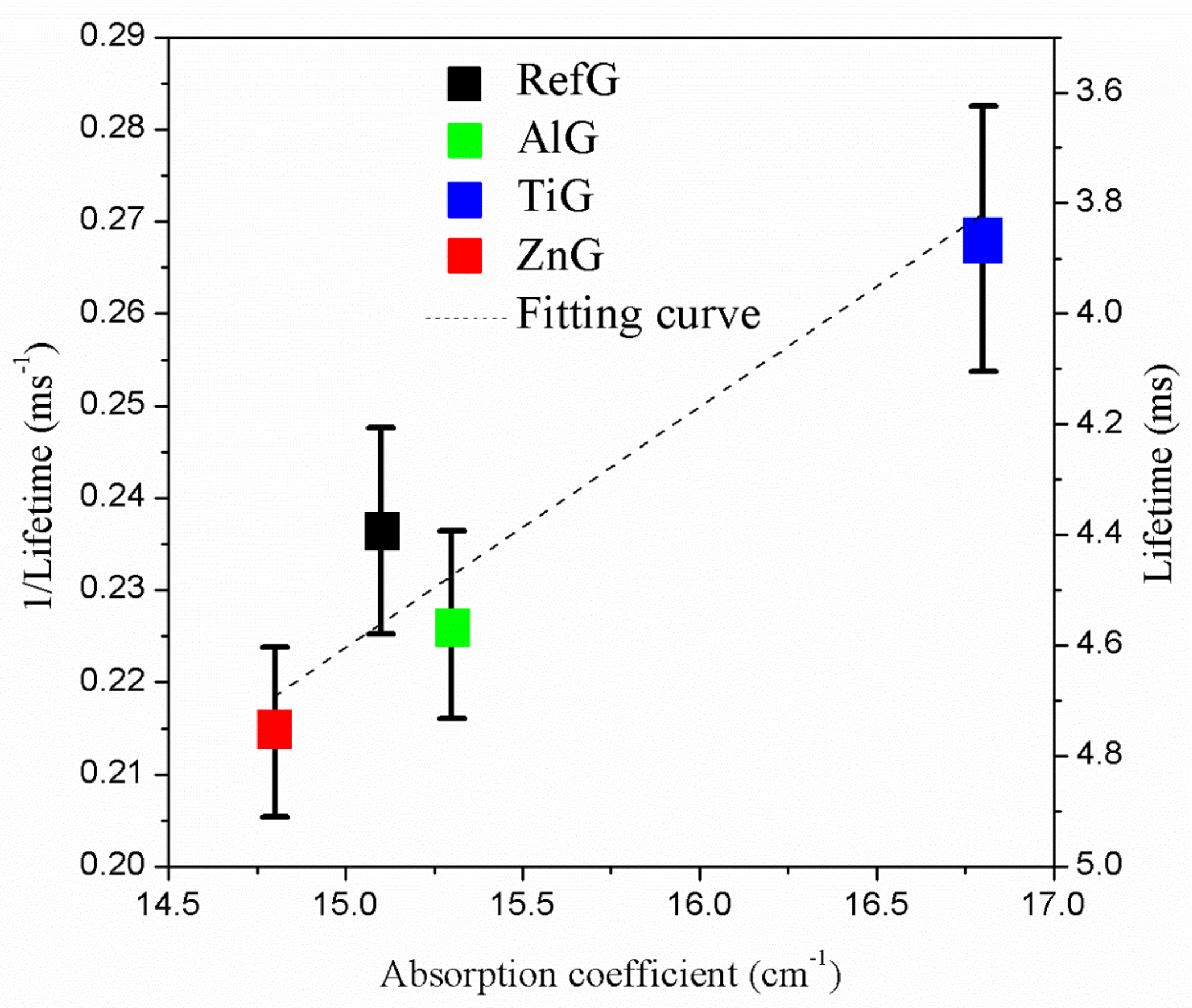

**Fig. 9.**

**Table 1** Physical, thermal and luminescence properties of the glasses.

| Glass label | $\rho$ ± 0.02 g/cm$^3$ | $T_g$ ± 3 °C | $T_p$ ± 3 °C | $\Delta T$ ± 6 °C | [$Er^{3+}$] (·$10^{19}$) (ions/cm$^3$) ± 5% | Abs. coeff. at 2900 cm$^{-1}$ related to [$OH^-$] | $\sigma_{Abs}$ at 1550 nm (·$10^{-21}$) (cm$^2$) ± 10% | $\sigma_{Abs}$ at 980 nm (·$10^{-21}$) (cm$^2$) ± 10% | $\sigma_e$ at 1550 nm (·$10^{-21}$) (cm$^2$) ± 10% | $Er^{3+}$:$^4I_{13/2}$ $\tau$ (ms) ± 0.20 ms |
|---|---|---|---|---|---|---|---|---|---|---|
| **RefG** | 3.08 | 440 | 555 | 115 | 7.775 | 15.1 | 6.08 | 1.89 | 5.78 | 4.23 |
| **AlG** | 3.10 | 447 | 583 | 136 | 7.843 | 15.3 | 5.02 | 1.79 | 4.85 | 4.42 |
| **TiG** | 3.09 | 448 | 563 | 115 | 7.840 | 16.8 | 6.25 | 2.04 | 6.01 | 3.73 |
| **ZnG** | 3.08 | 439 | 538 | 99 | 7.813 | 14.8 | 6.71 | 2.03 | 6.53 | 4.66 |